\documentclass[10pt,conference]{IEEEtran}
\usepackage{cite}
\usepackage{multirow}
\usepackage{amsmath,amssymb,amsfonts}
\usepackage{algorithm,algorithmicx,algpseudocode}
\usepackage{graphicx}
\usepackage{textcomp}
\usepackage{xcolor}
\usepackage[hyphens]{url}
\usepackage{fancyhdr}
\usepackage{comment}
\usepackage[hidelinks]{hyperref}
\usepackage{array}
\usepackage{pifont}

\title{DRACO: Co-design for DSP-Efficient Rigid Body Dynamics Accelerator\\
\thanks{* Corresponding author.}
}

\author{\IEEEauthorblockN{Xingyu Liu$^{1}$, Jiawei Liang$^{1}$, Yipu Zhang$^1$, Linfeng Du$^1$, Chaofang Ma$^1$, Hui Yu$^1$, Jiang Xu$^2$, Wei Zhang$^{1*}$}
\IEEEauthorblockA{$^1$\textit{The Hong Kong University of Science and Technology}, $^2$\textit{The Hong Kong University of Science and Technology (GZ)}\\
\{xliugu, jliangbr, yzhangqg, linfeng.du, cmaaw, reconcluster\}@connect.ust.hk, jiang.xu@hkust-gz.edu.cn, wei.zhang@ust.hk}
\vspace{-22pt}
}

\begin{document}
\maketitle


\begin{abstract}

Rigid Body Dynamics (RBD) computation is a critical component of robotic control, often dominating system runtime due to its algorithmic complexity and high parallelism demands. CPUs suffer from limited parallelism and cache-unfriendly access patterns, while GPUs incur prohibitive memory-access latency and per-task response time, making them unsuitable for real-time control. Both platforms also consume excessive power for edge deployment. FPGAs offer superior latency, energy efficiency, and customizable hardware-level parallelism, emerging as promising targets for RBD acceleration. However, existing FPGA designs still face critical limitations.

First, the intensive use of multiply-accumulate operations leads to high DSP consumption---especially for high degrees-of-freedom (DOF) robots---resulting in limited scalability. 
Second, RBD functions include mass matrix inversion function, which is inefficient on FPGA due to reciprocal operations falling on the longest latency path, severely limiting performance.
Third, mismatched processing rates across modules introduce idle cycles, resulting in poor DSP utilization.

To address these issues, we propose a hardware-efficient and high-performance RBD accelerator based on FPGA, introducing three key innovations.
First, we propose a precision-aware quantization framework that reduces DSP demand while preserving motion accuracy. This is also the first study to systematically evaluate quantization impact on robot control and motion for hardware acceleration.
Second, we leverage a division deferring optimization in mass matrix inversion algorithm, which decouples reciprocal operations from the longest latency path to improve the performance.
Finally, we present an inter-module DSP reuse methodology to improve DSP utilization and save DSP usage.
Experiment results show that our work achieves up to 8$\times$ throughput improvement and 7.4$\times$ latency reduction over state-of-the-art (SOTA) RBD accelerators across various robot types, demonstrating its effectiveness and scalability for high-DOF robotic systems.

\end{abstract}

\section{Introduction} \label{section_1}

\begin{figure} [t]
\centering
\includegraphics[width=\linewidth]{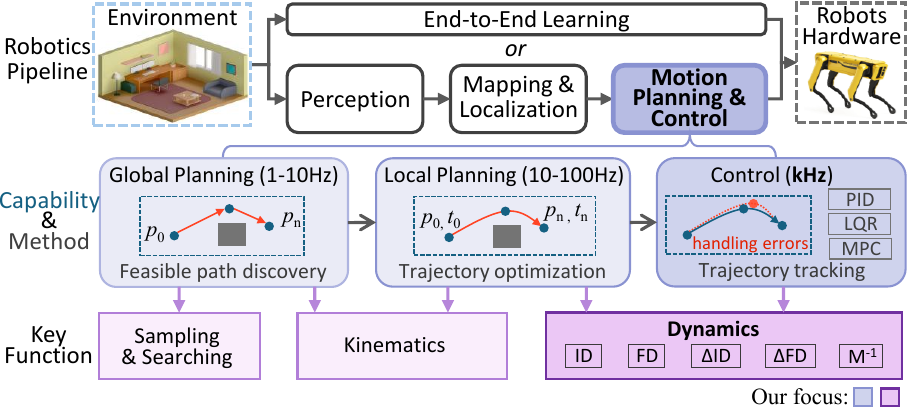}
\vspace{-20pt}
\caption{Robotics pipeline and motion planning \& control framework.}
\label{fig_robotics_introduction}
\vspace{-20pt}
\end{figure}

Robotic systems have evolved from simple manipulators to sophisticated autonomous agents capable of complex interactions with dynamic environments. As shown in Fig.~\ref{fig_robotics_introduction}, the robotics pipeline consists of three stages~\cite{wanSurveyFPGABasedRobotic2021,wanRoboticComputingFPGAs2022}: perception; mapping \& localization; and motion planning \& control. While end-to-end learning-based approaches~\cite{OpenVLAOpenSourceVisionLanguageAction2024} show promise in high-level decision making, they fall short under stringent real-time constraints of robotics and need to be combined with traditional motion planning \& control methods~\cite{Corki}. As a result, traditional methods remain the core solution for reliable low-level execution~\cite{DaduRBD}.

Among three stages of robotics pipeline, motion planning \& control is the most time-critical. It typically consists of three layers: global planning, local planning, and control~\cite{DaduRBD}, each operating at progressively higher frequencies and handling lower-level, more physically detailed tasks. Global planning computes feasible paths under geometric and kinematic constraints, while local planning refines these paths into executable trajectories under kinodynamic constraints. 

Control, operating at the highest rates, ensures precise trajectory tracking with dynamics compensation, translating planned motions into actuator commands while reacting to disturbances in real time. Common controllers include Proportional-Integral-Derivative (PID)~\cite{yuPIDControlIntelligent2018,wangDataDrivenOutputFeedbackFaultTolerant2016}, which applies real-time error correction; Linear Quadratic Regulator (LQR)~\cite{LQRcontrol}, generating optimal linear feedback gains to minimize a quadratic cost; and Model Predictive Control (MPC)~\cite{RealtimeMotionPlanningMPC,nguyenTinyMPCModelPredictiveControl2024}, that solves optimization problems over a prediction horizon for each control step.

\begin{figure*} [t]
\centering
\includegraphics[width=\linewidth]{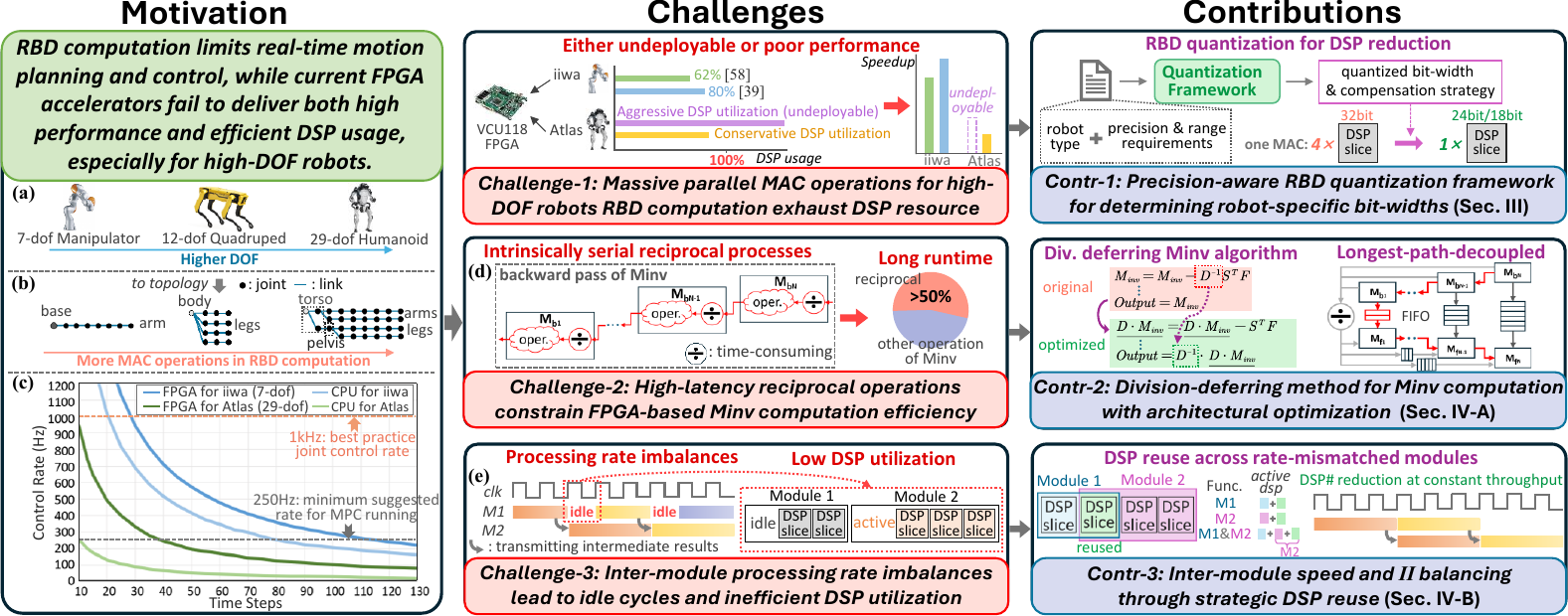}
\vspace{-20pt}
\caption{Overview of the motivation, challenges and contributions.}
\label{fig_motivation_challenge_contribution}
\vspace{-20pt}
\end{figure*}

While accelerators for planning meet performance requirements with low frequencies (1–100Hz), control operates at higher rates (up to kHz), posing stricter real-time constraints. At the heart of the controllers' computations lies rigid body dynamics (RBD), which models how joint torques, accelerations, and velocities interact during motion, based on robot’s structure. RBD includes a suite of functions implemented through recursive computation patterns and dense multiply-accumulate (MAC) operations. These functions often consume up to 90\% of controllers' total runtime~\cite{DaduRBD,AnalyticalDerivativesRSS,plancherPerformanceAnalysisParallel2020}, making RBD the primary computational bottleneck. 

RBD computation must meet two key performance requirements: low latency to ensure real-time control stability and high throughput~\cite{DaduRBD} to enable multi-time-step computations in long-horizon optimization methods. Existing approaches, however, face significant limitations. 

CPU-based solutions rely on multi-threading but suffer from cache-unfriendly computation patterns~\cite{DaduRBD} and low parallelism, leading to poor latency and throughput. GPUs~\cite{plancherAcceleratingRobotDynamics2021} offer higher throughput but struggle with high memory-access latency and large per-task response time. Both hardware platforms also consume considerable power~\cite{DaduRBD}, making them unsuitable for edge-deployed robotic systems.

FPGAs provide a compelling alternative, offering low latency, high parallelism, and high energy efficiency. Their reconfigurability also enables adaptation to diverse robot structures. Recent works thus explore FPGA-based accelerators for RBD computation. Robomorphic~\cite{Robomorphic} and Roboshape~\cite{Roboshape} leverage robot topology to design accelerators for specific RBD derivatives, while Dadu-RBD~\cite{DaduRBD} improves memory access efficiency, achieving high throughput across multiple RBD functions. Though promising, these designs still face key limitations. Roboshape suffers from limited throughput, and Dadu-RBD incurs high latency in mass matrix inversion. More broadly, none of existing works address the central issue: the large number of parallel MAC operations required by high degrees-of-freedom (DOF) robots quickly exhaust available DSP resources, leading to drastically degraded parallelism or even physical synthesis failure.

Fig.~\ref{fig_motivation_challenge_contribution}(a) and (b) show three typical robot topologies, and (c) highlights the performance gap in MPC implementations across robots with varying DOFs, comparing estimated control rates of the low-DOF iiwa~\cite{iiwa} and high-DOF Atlas~\cite{Atlas} under similar resource budget.
Even with FPGA acceleration, Atlas fails to maintain the ideal 1 kHz control rate required for direct joint-level MPC under long-horizon settings. Furthermore, despite similar DSP usage, Atlas exhibits significantly lower control rates than iiwa, highlighting the limited scalability of existing accelerators with increasing robot complexity.

To enhance performance and scalability of FPGA-based RBD accelerators, several challenges must be addressed:

\textbf{Challenge-1}: High-DOF robots require extensive parallel MAC operations for RBD computation, which quickly saturate DSP resources. Consuming 62\% and 80\% of VCU118 FPGA's DSPs, existing designs~\cite{DaduRBD,Roboshape} accelerate low-DOF iiwa effectively, but cannot scale to high-DOF robots like Atlas---either exceeding resource limits or sacrificing performance.

\textbf{Challenge-2}: The Mass Matrix Inversion (Minv) function is the major performance bottleneck in RBD functions, as it involves a series of reciprocal operations that are not efficiently supported by FPGA. These operations lie on the longest latency path (Fig.~\ref{fig_motivation_challenge_contribution}(d)), consuming over half of the runtime and significantly increasing latency.

\textbf{Challenge-3}: Variations in the computational loads of hardware modules induce mismatches in processing rates and initiation interval (\textit{II}), leading to poor DSP utilization. For example, as shown in Fig.~\ref{fig_motivation_challenge_contribution}(e), a simple module M1 completes its operations faster than a complex module M2, and DSPs in M1 instance are left idle in a significant portion of time. 

To overcome these challenges and meet the demand of real-time control, we propose DRACO, an FPGA-based hardware-efficient and high-performance RBD accelerator for robots with varying DOFs. Three pointed contributions to each challenge are summarized as follows: 

\begin{itemize}

\item We propose a quantization framework that automatically determines optimal bit-width configurations for RBD functions based on user-defined requirements, reducing DSP demand by up to 4$\times$ while maintaining motion accuracy. It is the first work conducting in-depth analysis and evaluation of the quantization impact on robotic control and motion for hardware acceleration.

\item We introduce a hardware-efficient Minv algorithm with division deferring optimization to remove reciprocal operations from the longest latency path in Minv, significantly improving performance and reducing overall latency.

\item We develop an inter-module DSP reuse methodology to address \textit{II} imbalances between hardware modules. By dynamically reallocating resources across modules, this approach improves DSP utilization, enabling substantial DSP savings and efficient scaling for high-DOF robotic systems while maintaining high performance.

\end{itemize}

DRACO achieves up to 8$\times$ throughput growth and 7.4$\times$ latency reduction compared to SOTA works~\cite{DaduRBD,Roboshape} across various robot types, while maintaining similar DSP usage.

\section{Background}

\subsection{Fundamentals of Robot Modeling}

Robots can be modeled in general formats~\cite{RBDAlgorithms,Robomorphic} that capture physical structure and dynamics, enabling consistent analysis across different systems. Topology trees, as shown in Fig.~\ref{fig_motivation_challenge_contribution}(a) and (b), are widely used to represent open-chain robots~\cite{RBDAlgorithms} with $N_B$ rigid links connected by $N_B$ joints. 

The i-th link’s mass and rotational inertia are represented by a symmetric inertia matrix $I_i \in \mathbb{R}^{6 \times 6}$, while i-th joint is characterized by a motion subspace matrix $S_i \in \mathbb{R}^{6 \times N_i}$, where $N_i$ is the joint’s DOF. For common joint types such as revolute and prismatic, $N_i = 1$ and $S_i$ is a one-hot vector. The spatial relationship between adjacent links is given by a transformation matrix $^iX_{\lambda_i}$, where $\lambda_i$ is the parent link’s ID.

Key parameters such as $I_i$, $S_i$, and calibrated values in $^iX_{\lambda_i}$ are constants for a given robot, while joint states ($q$, $\dot{q}$, $\ddot{q}$) dynamically influence $^iX_{\lambda_i}$ and the equations of motion. This distinction between constant and state-dependent variables is critical for efficient computation in control.

\subsection{Rigid Body Dynamics Functions}

The foundation of robot rigid body dynamics lies in the equation of motion for a rigid body system:
\begin{equation} \label{equation_of_motion}
M(q)\ddot{q} + C(q, \dot{q}, f^{ext}) = \tau,
\end{equation}
where $q$, $\dot{q}$, $\ddot{q}$, and $\tau$ represent vectors of joint positions, velocities, accelerations, and forces/torques, respectively. \(M(q)\) is the symmetric positive definite mass matrix (also called joint space inertia matrix), and \(C(q, \dot{q}, f^{ext})\) denotes the generalized bias forces, which account for Coriolis, centrifugal, gravitational, and external forces not included in \(\tau\).

\begin{figure} [t]
\centering
\includegraphics[width=\linewidth]{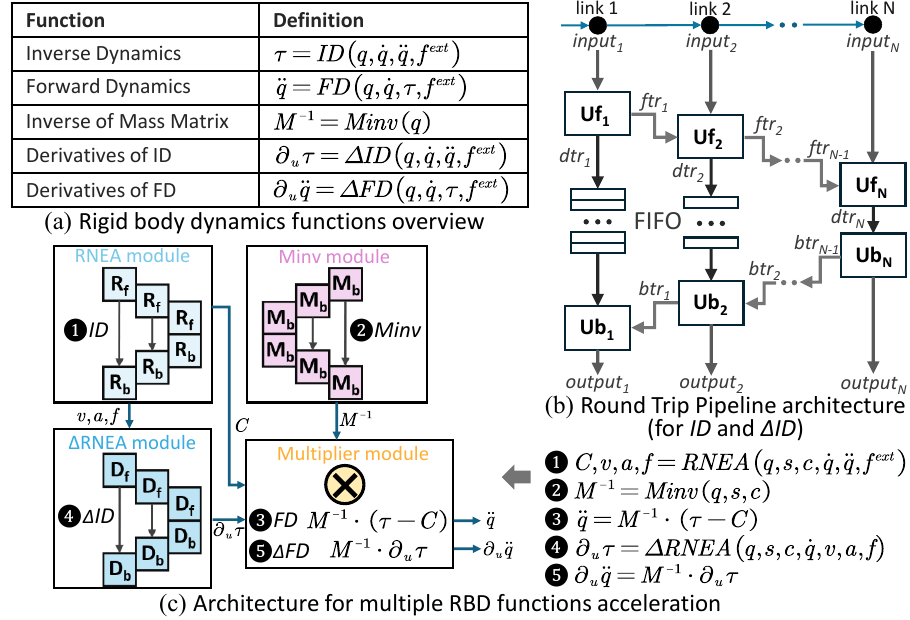}
\vspace{-22pt}
\caption{RBD functions and current architecture. (a) RBD functions overview. (b) Round Trip Pipeline architecture in~\cite{DaduRBD}. (c) Architecture for accelerating RBD functions.}
\label{fig_background}
\vspace{-19pt}
\end{figure}

From Eq.~\ref{equation_of_motion}, several RBD fundamental functions are derived~\cite{RBDAlgorithms,AnalyticalDerivativesRSS,DaduRBD,Robomorphic}, as summarized in Fig.~\ref{fig_background}(a). \textit{Inverse Dynamics} (ID) is typically computed using the Recursive Newton-Euler Algorithm (RNEA)~\cite{RBDAlgorithms}, and its derivatives ($\Delta$ID) are calculated using \(\Delta\)RNEA method~\cite{AnalyticalDerivativesRSS,plancherGRiDGPUAcceleratedRigid2022}. \textit{Forward Dynamics} (FD) and its derivatives are computed using:
\begin{equation} \label{equation_fd}
FD = M^{-1}ID, \quad \Delta FD = M^{-1}\Delta ID.
\end{equation}
The inverse of the mass matrix \((M^{-1})\) is efficiently computed using the algorithm in ~\cite{carpentierAnalyticalInverseJoint} (called Minv algorithm in this paper), which will be elaborated in Sec.~\ref{division_defer_Minv}.

A notable feature of these dynamics functions is their reliance on \textbf{bidirectional traversals} (forward and backward propagation)~\cite{DaduRBD,Robomorphic}, reflecting the chain-like dependencies of robot joints. Both RNEA and \(\Delta\)RNEA involve a forward pass (base to end-effector~\cite{oltjenReductionEndEffector2015}) followed by a backward pass (end-effector to base). Intermediate results from the forward pass are reused during the backward pass. This bidirectional dataflow leads to poor temporal locality, which subsequently increases cache misses and memory access, posing challenges for CPUs and GPUs~\cite{DaduRBD}. FPGAs, with their flexible memory configurations, are well-suited to mitigate these limitations and accelerate RBD computations.

\subsection{Current Accelerator Architectures}

To address the computational demands of dynamics functions, Robomorphic~\cite{Robomorphic} implements separate computing cores for forward and backward passes, supported by schedulers and global buffers to store intermediate results. In contrast, Dadu-RBD introduces a Round Trip Pipeline (RTP) architecture optimized for FPGAs, achieving higher throughput via massively pipelined parallelism, as illustrated in Fig.~\ref{fig_background}(b). Each pipeline stage (Uf/Ub) corresponds to a joint, with FIFO buffers temporarily storing intermediate results between the forward and backward passes for the same joint. The FIFO buffers act as pipeline registers, ensuring seamless dataflow and eliminating the need for additional memory access. Different RTP modules are combined into a multi-function architecture (Fig.~\ref{fig_background}(c)), enabling Dadu-RBD to efficiently support diverse RBD computations with flexibility. Our work builds on the Dadu-RBD architecture, extending it with targeted optimizations and enhancements.

\section{Precision-Aware Quantization Methodology}

\subsection{Motivations, Insights, and Challenges of RBD Quantization}

\textbf{Motivations.} Quantization is a widely adopted technique to reduce computational cost and save hardware resources, particularly in matrix-heavy computations like neural networks. Its advantages are promising for accelerating RBD, where using lower-bit fixed-point operations can drastically reduce resource usage. For instance, on FPGAs, a 32-bit MAC consumes four DSP48 slices, while an 18-bit MAC typically uses only one~\cite{DSP48}.

\textbf{Insights.} Quantization inevitably introduces numerical errors, raising concerns for its applicability in high-precision robotic control domain.
RBD computations are closely associated with control accuracy and stability, and small numerical deviations can, in theory, affect real-time motion behavior. However, prior work~\cite{hsiaoVaPrVariablePrecisionTensors2023} and real-world observations suggest that robotic systems typically exhibit a degree of error tolerance.
For example, ideal rigid body assumptions rarely hold true in actual systems and joints often experience friction and compliance~\cite{RBDAlgorithms}. Yet robots still perform effectively under such conditions. This observed robustness inspires the use of quantization for hardware-efficiency gains. The key challenge lies in ensuring that quantization-induced errors remain within acceptable bounds.

Despite its promise, quantization in robotic control and RBD remains underexplored. Existing RBD implementations are limited to floating-point arithmetic~\cite{PinocchioLibrary} or 32-bit fixed-point formats (16 int / 16 frac)~\cite{DaduRBD,Robomorphic,Roboshape}. While easier to implement, these formats fail to achieve optimal resource efficiency. More critically, these works rarely evaluate quantization effects on control performance. For instance, Robomorphic~\cite{Robomorphic} evaluates MPC controllers under different bit-width configurations but focuses solely on optimization loss, which does not directly reflect motion accuracy. Its analysis is restricted to 32-bit fixed-point formats and limited to MPC, leaving open questions regarding other bit-widths and controller types. These limitations highlight the need for a comprehensive, precision-aware quantization framework that supports various controllers and adapts to diverse applications—an objective we target in this work.

\textbf{Challenges.} Quantizing RBD algorithms presents distinctive challenges that set them apart from other domains:
\begin{itemize}
    \item Unpredictable error behavior. Quantization errors in RBD computations propagate in highly nonlinear and input-dependent ways~\cite{RBDAlgorithms}, influenced by robot type, application scenario, and control strategy. This unpredictability complicates analytical modeling and makes it difficult to generalize error behavior.
    \item Diverse precision and value range requirements. Robotic systems vary widely in structure, actuation, and application—resulting in diverse precision and value range requirements for RBD computation. For instance, some robotic applications need strict accuracy, while some can tolerate coarser quantization. This variability necessitates a general quantization methodology that searches for optimal bit-width configurations under varying requirements, balancing control precision and hardware efficiency.
\item Controller-specific precision sensitivity.  
Different control strategies exhibit varying sensitivity to RBD quantization. PID is highly sensitive due to reliance on accurate dynamics for real-time compensation~\cite{yuPIDControlIntelligent2018}. LQR, though less sensitive, still depends on faithful RBD-based linear approximations. MPC embeds RBD within its optimization loop and benefits from iterative correction, making it more tolerant to quantization noise. These differences necessitate controller-aware quantization strategies, as further examined in our evaluation.
\end{itemize}

\begin{figure} [b]
    \centering
    \vspace{-15pt}
    \includegraphics[width=\linewidth]{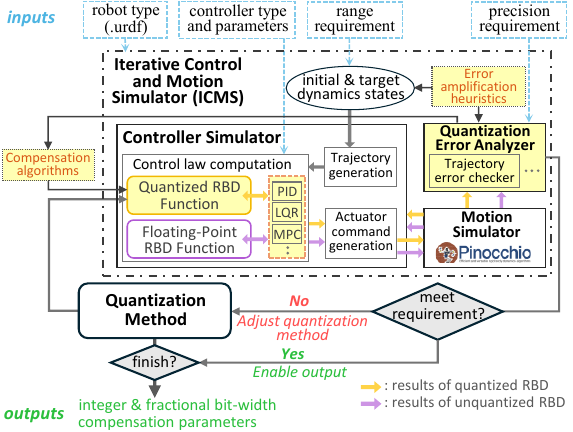}
    \vspace{-20pt}
    \caption{Overview of the quantization framework.}
    \label{fig_quantization_framework}
\end{figure}

\begin{figure*} [t]
\centering
\includegraphics[width=\linewidth]{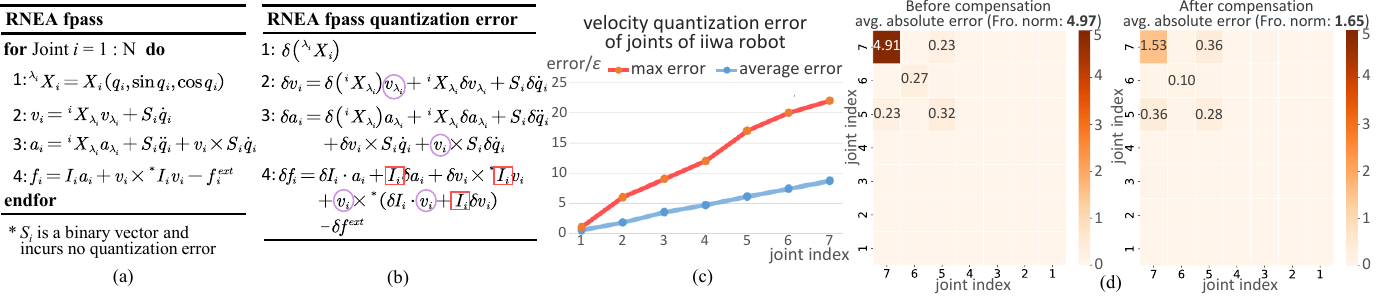}
\vspace{-22pt}
\caption{Error propagation and error compensation. (a) Forward pass of RNEA algorithm. (b) Quantization error of RNEA's fpass. (c) Velocity quantization error of joints of iiwa robot. (d) Average absolute error of quantized $M^{-1}$ matrix before and after compensation. (Small errors are not shown in this figure.)}
\label{fig_error_propagation}
\vspace{-18pt}
\end{figure*}

\subsection{Precision-Aware Quantization Framework} \label{Sec_Precision_Aware_Quantization_Framework}

To address the challenges of RBD quantization, we propose a precision-aware fixed-point quantization framework (Fig.~\ref{fig_quantization_framework}). The framework stands out for its adaptability to diverse controllers, systematic error evaluation, and uniform bit-width configuration to improve computation efficiency while maintaining motion accuracy across various robotic applications. It combines mathematical modeling and simulation-based error evaluation to search for optimal fixed-point formats, supported by heuristic error amplification patterns that accelerate the search and compensation algorithms which effectively mitigate quantization errors during deployment.

\textbf{Inputs.} Users provide robot's \texttt{urdf} description~\cite{URDF}, from which our framework extracts relevant properties like topology and inertia. Control strategy (PID, LQR, or MPC) with configurations is also specified to tailor quantization methods for controller-specific requirements. Optional inputs include variable range constraints (e.g., joint position, velocity limits). Precision requirements for motion accuracy are defined by users, such as trajectory error tolerances and error bounds of physical quantities.

\textbf{Framework Workflow.}  
As shown in Fig.~\ref{fig_quantization_framework}, the core component of quantization framework is the Iterative Control and Motion Simulator (ICMS), which performs closed-loop control simulation to model how quantization affects motion and control outcomes. Rather than relying purely on theoretical error bounds or black-box learning~\cite{Tartan}, ICMS enables reliable precision validation under realistic dynamic interactions—capturing both control behavior and physical response, and providing feedback to guide quantization choices. 

The simulation loop begins by feeding dynamics state samples—either randomly generated or selected from real datasets—into the controller, in which three control templates (PID, LQR, MPC) are pre-implemented. These templates are configurable: users can select the type, set key parameters, and adapt dynamics formulations. The controller computes both floating-point and quantized versions of RBD functions. Control outputs are passed into a Motion Simulator, built upon the Pinocchio RBD library~\cite{PinocchioLibrary}, to compute updated joint states and robot motion. This process iterates to form a closed-loop system that reflects how quantization affects both control response and robot motion.

A key innovation of the framework is the Quantization Error Analyzer in ICMS. It incorporates multiple error propagation and amplification heuristics derived from mathematical analysis to accelerate convergence and eliminate poor format candidates early in the search (detailed in Sec.~\ref{sec_error_compensation}). This strategy allows the framework to prune low-performing candidates without running full simulations across all configurations. To support varied application needs, the analyzer can evaluate multiple precision metrics—like trajectory deviation, posture error, and control torque error—which can be selected or weighted based on task requirements.

The framework also includes compensation algorithms to mitigate quantization error. These algorithms model common error structures and apply parameterized corrections which are computed and fine-tuned within the simulation loop (detailed in Sec.~\ref{sec_error_compensation}).

\begin{figure*} [t]
\centering
\includegraphics[width=\linewidth]{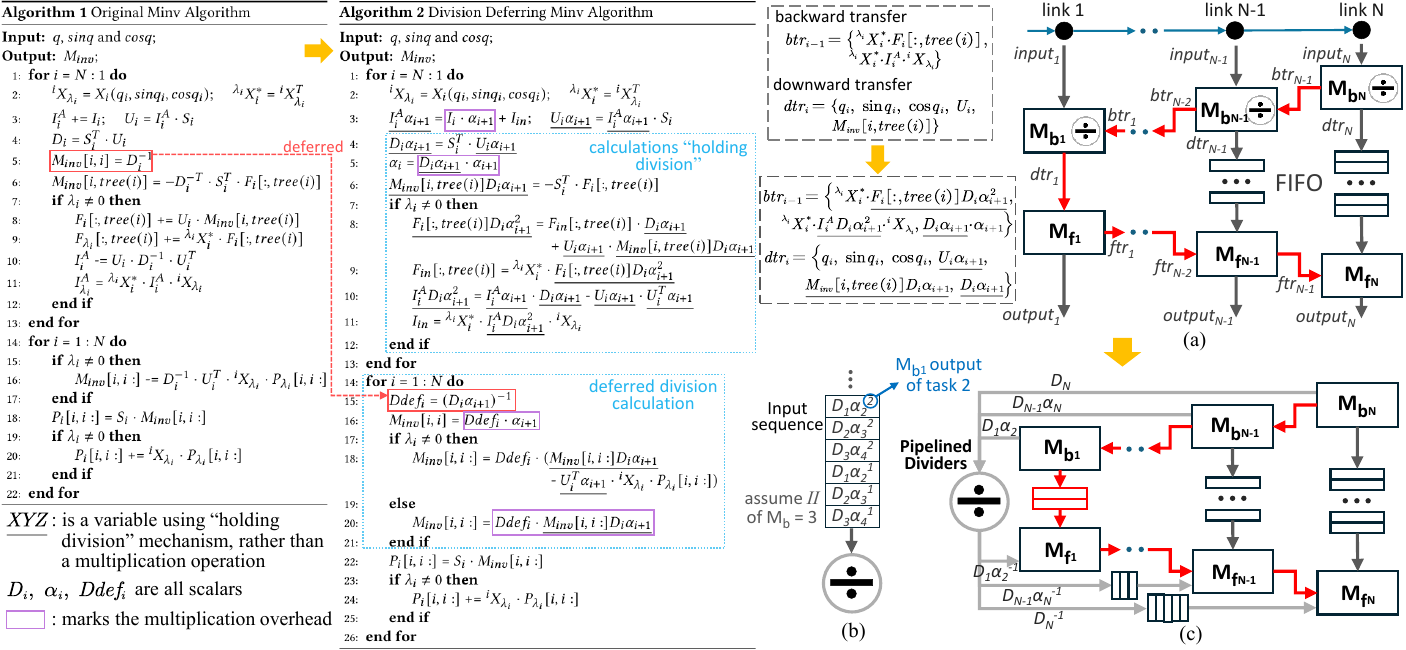}
\vspace{-22pt}
\caption{Minv algorithms and architectures before and after optimization. (a) Original architecture for Minv computation. (b) Example of the divider input sequence. (c) Optimized architecture for division deferring Minv computation.}
\label{fig_division_deferring}
\vspace{-18pt}
\end{figure*}

\textbf{Outputs.} The framework outputs an optimal fixed-point format with a uniform integer and fractional bit-width across all RBD variables. For FPGAs, 18-bit and 24-bit formats are prioritized to match DSP input bit-widths~\cite{DSP48,DSP58}, with sub-18 and mid-range widths (19–23) excluded due to few DSP resources saving. If 24-bit fails to meet the requirements, higher precisions are explored.  Final outputs also include error compensation parameters for hardware implementation.

\textbf{Key Properties.} This framework offers a general and controller-adaptive path toward quantizing RBD computations while preserving motion accuracy. Its integration of simulation-based validation, heuristic-driven candidate pruning, and compensation mechanisms makes it applicable across diverse controllers and robotic platforms. The unified bit-width configuration further simplifies hardware implementation and aligns well with standard DSP word sizes in FPGA designs.

Beyond FPGAs, the framework can also be adapted for ASIC-based RBD accelerators~\cite{Robomorphic}. In ASIC designs, quantization is crucial for reducing chip area and power—key factors for embedded and edge scenarios. Without the bit-width constraints of DSP blocks, our framework supports finer-grained bit-width searches, trading search time for improved silicon efficiency. Although this work focuses on uniform bit-width quantization, future exploration of mixed-precision schemes tailored to specific RBD functions holds promise, especially in the context of ASIC accelerators. By bridging control accuracy with hardware efficiency, the framework establishes a principled foundation for quantizing RBD computations across diverse robotic systems and hardware platforms.

\subsection{Error Amplification and Compensation} \label{sec_error_compensation}

To guide quantization search and mitigation precision loss, our framework incorporates mathematical error propagation analysis and Monte Carlo experiments. While detailed derivations and experiment designs are omitted due to space constraints, we summarize the key insights and heuristics embedded in the framework.

Taking RNEA's forward pass as an example. Fig.~\ref{fig_error_propagation}(a) illustrates the computational process, and Fig.~\ref{fig_error_propagation}(b) shows the derived quantization error expression. For a single quantized variable, the error is bounded by:
\begin{equation} \label{equation_of_quantization}
\left| \delta x \right|=\left| x-\frac{round\left( x\cdot 2^{n_{frac}} \right)}{2^{n_{frac}}} \right|\le 2^{-n_{frac}-1}=\varepsilon,
\end{equation} 
where $\delta x$ denotes the error, and $n_{frac}$ is fractional bit-width. 

While variables like $^iX_{\lambda_i}$ (transformation matrix) exhibit minimal precision loss from direct quantization, propagation of errors through dependent computations significantly increases complexity. For instance, the error in the force output $f_i$ from RNEA's forward pass is given by (Fig.~\ref{fig_error_propagation}(b), line 4):
\begin{equation}
\begin{aligned}
\delta f_i=&\delta I_i\cdot a_i+I_i\delta a_i+\delta v_i\times ^*I_iv_i \\
&+v_i\times ^*\left( \delta I_i\cdot v_i+I_i\delta v_i \right) -\delta f^{ext}
\end{aligned}
\end{equation} 
This expression demonstrates that once multiple quantized variables interact, error propagation becomes highly entangled, resulting in complex coupling patterns that are difficult to analyze analytically. To address this, our framework incorporates \textbf{heuristic error amplification patterns guiding the quantization search}—derived from theoretical analysis and validated through simulation—to prioritize critical variables for evaluation and prune unsuitable precision formats early in the search process.

\ding{202} Joint-depth accumulation. Errors from joints closer to the base accumulate as they propagate toward the end-effector, increasing total error with depth (Fig.~\ref{fig_error_propagation}(c)). Using this characteristic, the Quantization Error Analyzer prioritizes deeper joints for evaluation to reduce ineffective candidate testing.

\ding{203} Inertia-induced amplification. Large values in inertia matrix $I_i$ (box in Fig.~\ref{fig_error_propagation}(b)) amplify quantization errors when multiplied by error terms. Using this feature, joints with larger $I_i$ values are prioritized for evaluation to quickly eliminate incompatible quantization methods.
    
\ding{204} High-speed amplification. High-speed states amplify quantization noise in velocity-dependent terms (circle in Fig.~\ref{fig_error_propagation}(b)). These states are evaluated first in simulation to reject underperforming quantization formats.

\textbf{Error Compensation.}  
To mitigate quantization errors in RBD functions, the framework incorporates heuristic compensation strategies in quantized RBD computing, as illustrated in Fig.~\ref{fig_quantization_framework}. These strategies are derived from mathematical analysis and applied to components most affected by quantization yet relatively insensitive to dynamic input variations. For example, fixed-pattern corrections are used in operations like matrix inversion, where numerical distortion is primarily structural rather than trajectory-dependent. While the compensation algorithms are general-purpose and fixed within the framework, the actual parameters of these algorithms are tailored to each robot and application. During iterative simulation, these parameters are adjusted based on robot topology and statistical analysis of error behavior in quantization error analyzer. Final parameters are exported for RTL-level integration into accelerator. 

A representative case is the Minv algorithm (Alg. 1, Fig.~\ref{fig_division_deferring}), where a customized offset matrix is applied to the quantized \( M^{-1} \)to counter significant errors from reciprocal operations. This compensation primarily targets the diagonal terms, which are the main source of error propagation. Since off-diagonal elements are computed from these terms, the targeted correction can have a minor, non-uniform effect on them, as seen by a slight error increase from 0.23 to 0.36 in Fig.~\ref{fig_error_propagation}(d). However, this trade-off proves highly effective. The overall error, quantified by the Frobenius norm~\cite{FrobeniusNorm}, is drastically reduced from 4.97 to 1.65, demonstrating the robustness and success of our compensation strategy.

\begin{figure*} [t]
    \centering
    \includegraphics[width=\linewidth]{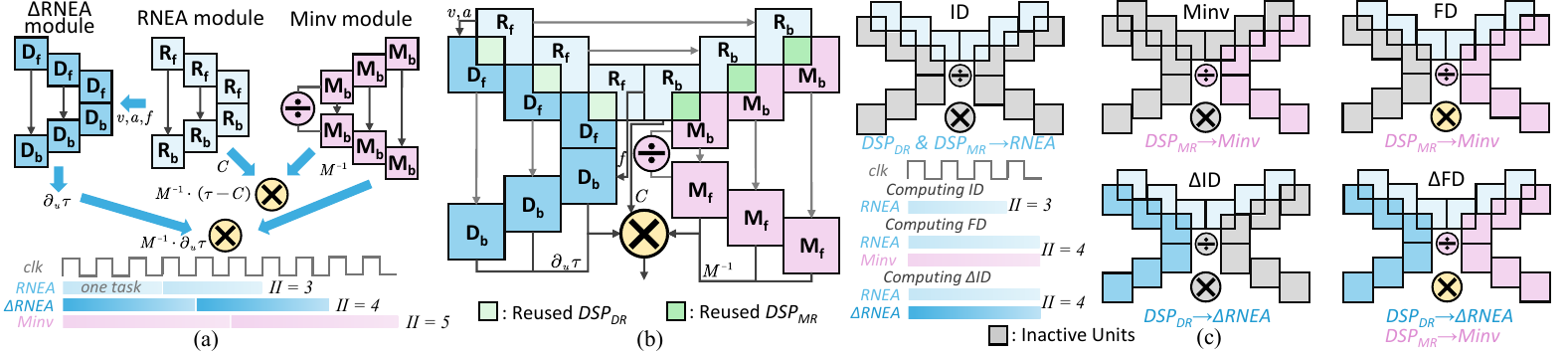}
    \vspace{-22pt}
    \caption{Inter-Module pipeline imbalance and proposed DSP reuse strategy. (a) Example of unbalanced computation speed between different basic modules. (b) A simplified diagram of the DRACO architecture. (c) Modules activated when accelerating various functions}
    \label{fig_inter_module_dsp_reuse}
    \vspace{-18pt}
\end{figure*}

\section{Architecture Optimization}

This section introduces two complementary optimization techniques aimed at improving the performance and resource efficiency of RBD accelerators. The first focuses on removing bottlenecks in Minv computation, while the second applies resource sharing across modules to maximize hardware utilization. Together, these techniques provide a holistic approach to optimizing RBD computations for high-DOF and resource-constrained robotic systems.

\subsection{Hardware-Efficient Minv Algorithm via Division Deferring} \label{division_defer_Minv}

The Minv function is a critical component in the computation of FD and $\Delta$FD (Fig.~\ref{fig_background}), significantly impacting performance. As shown in Fig.~\ref{fig_latency_throughput}, the latency and throughput of Minv function implemented by Dadu-RBD~\cite{DaduRBD} are comparable to those of FD, indicating that the overall performance of FD heavily depends on the efficiency of Minv computation.  

The original Minv algorithm~\cite{carpentierAnalyticalInverseJoint}, shown in Algorithm 1 of Fig.~\ref{fig_division_deferring}, performs a reciprocal operation (line 5) during the backward propagation. This operation lies on Minv’s longest latency path (Fig.~\ref{fig_division_deferring}(a)) and accounts for over 50\% of the runtime, as detailed in \textbf{Challenge-2} (Sec.~\ref{section_1}). Since FPGAs are inefficient at performing fixed-point division (e.g., 32-bit division at 200 MHz requires 20 clock cycles), Dadu-RBD~\cite{DaduRBD} converts fixed-point numbers to floating-point for division~\cite{Floating-pointv7.1}, and then converts the results back to fixed-point. However, this approach increases hardware resource usage and power consumption, with little performance improvement.

To address this, we propose a hardware-efficient reformulation of the Minv algorithm (Algorithm 2 in Fig.~\ref{fig_division_deferring}), incorporating a technique we refer to as \textbf{division deferring}. The key idea is to remove the reciprocal operation from the longest latency path by deferring its dependent computations, thereby enabling division to execute in parallel with other operations. Specifically, in backward pass, computations involving \( D_i^{-1} \) are postponed using a “holding division” mechanism. Instead of executing division inline within the backward pass, both sides of the equation are multiplied by \( D_i^{-1} \) or other \textit{holding factors}, and intermediate results are propagated to the next backward unit (Mb). This introduces a transfer coefficient \( \alpha \) (line 5 of Algorithm 2), which accumulates the effects of successive multiplications caused by holding division. These coefficients are resolved during the forward pass, where division is performed to produce the final result. By deferring the use of the division result to the forward pass, the reciprocal operation is decoupled from the longest latency path.

The proposed algorithm is implemented with a new architecture (Fig.~\ref{fig_division_deferring}(c)). Division module is removed from Mb units; instead, intermediate results \( D_{i-1}\alpha_i \) from Mb are fed into a shared divider, and division results are passed to forward units (Mf). Division executes in parallel with other computations. 

A key feature of this architecture is the staggered scheduling of \( D_{i-1}\alpha_i \) outputs from multiple Mb units, ensuring continuous utilization of the fully pipelined divider and avoiding idle cycles. For example, as shown in Fig.~\ref{fig_division_deferring}(b), with an \textit{II} of 3 cycles for Mb, three Mb units can share a single divider instead of requiring one divider per unit. This optimization reduces the number of required dividers and lowers resource usage. Additionally, using a fully pipelined divider increases the circuit's clock frequency~\cite{DividerGeneratorv5.1}, despite minor latency from added pipeline stages.

The hardware-efficient Minv algorithm introduces modest resource overheads to achieve these performance gains. First, as highlighted in the purple box in Algorithm 2, it adds scalar and sparse matrix–scalar multiplications, incurring minimal DSP overhead within resource budget. Second, deferring division results leads to extra data transfers between Mb and Mf units, increasing FIFO usage. However, as LUTs and FFs are not resource-bond, this overhead is negligible. Finally, inserting additional FIFO buffer between Mb1 and Mf1 is required to temporarily store intermediate results before division completes. Although this introduces a few cycles of latency, it is far outweighed by overall latency reduction from decoupling division.

\subsection{Inter-Module DSP Reuse} \label{inter_module_dsp_reuse}

\begin{figure*} [t]
    \centering
    \includegraphics[width=\linewidth]{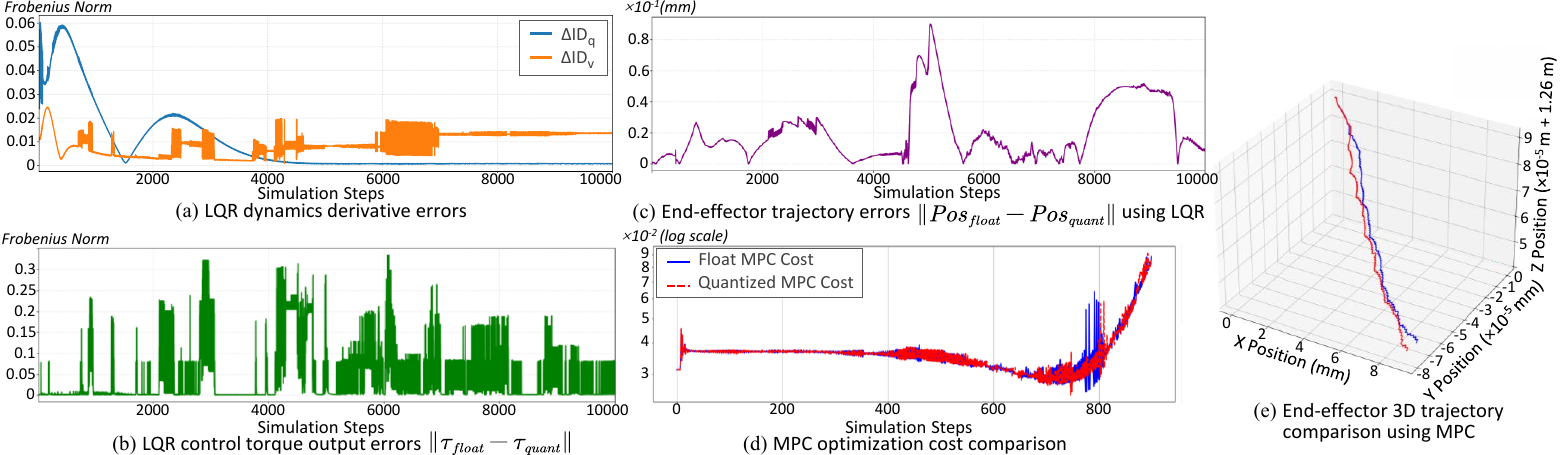}
    \vspace{-22pt}
    \caption{(a) LQR dynamics derivative errors, (b) LQR control torque output difference, (c) end-effector trajectory error difference using LQR, (d) MPC optimization cost comparison, and (e) end-effector 3D trajectory comparison using MPC for iiwa robots.}
    \label{fig_PID_LQR_MPC}
    \vspace{-18pt}
\end{figure*}

Building on the RTP architecture from Dadu-RBD and our optimized \( \text{Minv} \) algorithm, the internal performance of $basic$ $modules$ (Fig.~\ref{fig_inter_module_dsp_reuse}(a)) like RNEA, \( \text{Minv} \), and \( \Delta\text{RNEA} \) have been improved, significantly accelerating the computation of individual RBD functions. However, complex functions like FD, \( \Delta\text{ID} \), and \( \Delta\text{FD} \) (use results from basic modules as inputs), which require multiple basic modules to work in coordination (Fig.~\ref{fig_background}(c)), still suffer from imbalanced processing rates across modules. This issue leads to resource underutilization when faster modules are forced to wait for slower ones. For instance, as shown in Fig.~\ref{fig_inter_module_dsp_reuse}(a), RNEA and \( \Delta\text{RNEA} \) have different \textit{II}s. After sending intermediate results to \( \Delta\text{RNEA} \) module, RNEA module must stall for a cycle to align with the slower \( \Delta\text{RNEA} \), introducing idle cycles.

In addition to inter-module imbalances, processing rates imbalances also exit within individual basic modules~\cite{DaduRBD}. For example, in \( \Delta\text{RNEA} \) module, units (Df/Db) closer to the end-effector handle heavier computational loads~\cite{Roboshape}. Prior work~\cite{DaduRBD} addressed this by allocating more DSPs to high-load units and reusing DSPs in lighter-load ones, ensuring pipeline efficiency between units inside each basic module. Inspired by this approach, we propose inter-module DSP reuse to balance \textit{II} across modules, improve hardware utilization, and free DSPs for other computations.

Leveraging insights into functional dependencies and computational demand, we design the DRACO architecture, shown in Fig.~\ref{fig_inter_module_dsp_reuse}(b).
In this architecture, DSPs are shared between pairs of collaborating modules:
RNEA and \( \Delta\text{RNEA} \), and RNEA and \( \text{Minv} \).
As shown in Fig.~\ref{fig_inter_module_dsp_reuse}(c), resource activation dynamically changes based on the computed RBD function.

When only ID is calculated, shared DSP groups (\( DSP_{DR} \), \( DSP_{MR} \)) are fully allocated to the RNEA module (Fig.~\ref{fig_inter_module_dsp_reuse}(c), upper left) for maximum performance. When \( \text{Minv} \) is computed independently, \( DSP_{MR} \) is allocated to \( \text{Minv} \) module (Fig.~\ref{fig_inter_module_dsp_reuse}(c), upper middle). 

When RNEA module collaborates with other high-demand modules, the shared DSPs are dynamically reallocated to these modules to balance \textit{II}. For computing FD, \( \Delta\text{ID} \), and \( \Delta\text{FD} \), RNEA forgoes shared DSPs entirely, allowing \( \text{Minv} \) and \( \Delta\text{RNEA} \) modules to use these resources exclusively and meet their performance needs. This adaptive sharing ensures inter-module \textit{II} alignment (Fig.~\ref{fig_inter_module_dsp_reuse}(c), lower left), eliminating idle cycles and improving hardware utilization without degrading computation throughput. While it introduces minor latency overheads due to dynamic DSP sharing, the DSP savings allow reallocation to high-demand modules, significantly improving overall performance.

The inter-module DSP reuse technique can serve as a practical design pattern for future automatic RBD accelerator generation~\cite{Robomorphic,Roboshape}, especially for high-DOF robotic systems. To integrate such reuse into automated hardware generation flows, we propose three practical guidelines: \textbf{First}, configure reused DSPs count according to the \textit{II} mismatch between modules—larger mismatches demand more resource sharing. \textbf{Second}, adapt DSP reuse quantities to per-joint computational demands, as units near the end-effector often incur heavier MAC loads. \textbf{Third}, prioritize reuse across MAC operations with matched input dimensions or operand similarity, minimizing control overhead and simplifying datapath multiplexing.

Although implemented on FPGAs, our DSP reuse approach can be naturally applied to ASIC-based designs, where shared fixed-point multipliers can be allocated using similar principles. This extends the method’s applicability to a broader set of RBD accelerators, enabling efficient arithmetic resource reuse across pipeline stages and modules.

\section{Evaluation}

We first evaluate the quantization framework, show quantization's impact on control and motion, and give quantization methods for RBD functions of different robots based on typical precision and range requirements. Then we assess the performance, resource usage and power consumption of our FPGA accelerator. Finally, we estimate the control rate improvements achieved by our design.

\subsection{Quantization Framework Evaluation}

\begin{figure*} [t]
    \centering
    \includegraphics[width=170mm]{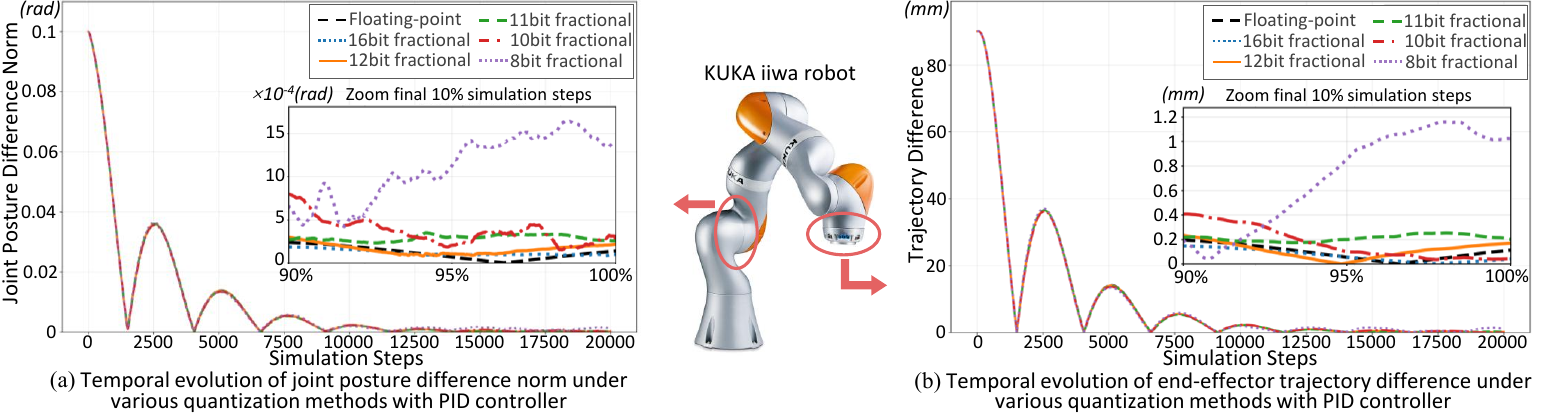}
    \vspace{-10pt}
    \caption{(a) Temporal evolution of the iiwa's second joint posture difference (distance to the target posture) norm under various quantization methods. (b) Temporal evolution of the iiwa's end-effector trajectory difference in Cartesian Space under various quantization methods.}
    \label{fig_PID}
    \vspace{-18pt}
\end{figure*}

\textbf{Motion Precision Metrics.}
Previous works evaluate performance using task-level success rates~\cite{Corki,hsiaoVaPrVariablePrecisionTensors2023}, which measure task completion under specific conditions. However, such metrics are influenced by task tolerances, environmental factors, and dataset biases. For instance, Corki~\cite{Corki} reports high success rates despite trajectory errors exceeding 1mm, illustrating the disconnect between binary task outcomes and actual motion precision.

\textbf{Trajectory Error as Evaluation Metric.}
While ISO metrics~\cite{ISOManipulate} describe static precision and success rate reflects high-level outcomes, neither captures deviations during continuous motion. Thus we adopt trajectory error as our evaluation metric, extending static metrics by measuring real-time deviations throughout the motion trajectory. Unlike task success rates, trajectory error directly reflects motion accuracy without being masked by task-specific tolerances. This makes it particularly suitable for evaluating quantization effects, as it isolates the impact on motion quality from application-specific factors.

For our evaluation, we focus on the iiwa robot due to its strict precision requirements, using a ±0.5mm trajectory error tolerance, much more stringent than ~\cite{Corki}. This sub-millimeter focus ensures quantization method meets the demands of high-precision applications.

For dynamic robots like HyQ and Atlas, which prioritize mobility and agility, trajectory error tolerance is relaxed to reflect broader motion constraints, maintaining fair evaluation across platforms.

\textbf{Quantization Effect Evaluation.} We evaluate the impact of quantization on three controllers: LQR, MPC, and PID with dynamic compensation. To isolate quantization effects, controller settings are kept simple and conventional, deliberately avoiding robust tuning, which could mask quantization sensitivity and hinder accurate assessment. In practical deployments, combining the framework with robust controllers would enable more aggressive quantization and further improve hardware efficiency. For iiwa, our framework searches optimal quantization formats for each controller without considering FPGA bit-width constraints. The obtained configurations use 12-bit integer and 12-bit fractional for PID, 10-bit for LQR, and 9-bit for MPC. These controller-specific formats are used in subsequent evaluations.

Fig.~\ref{fig_PID_LQR_MPC}(a)-(c) show, for a specific case, the dynamics derivative errors after quantization, the norm difference of control torque, and end-effector trajectory error, respectively, under LQR control. These three metrics reflect quantization-induced deviations at different stages: RBD function output, controller output, and final robot trajectory (our precision metric). LQR, which minimizes a cost function to derive the optimal control law, exhibits limited sensitivity to quantization errors in dynamics derivatives. As a result, deviations in control torque and trajectory error (less than 0.01mm) remain negligible. Similarly, the MPC controller shows less than 0.02mm trajectory deviations (Fig.~\ref{fig_PID_LQR_MPC}(e)) despite visible effects on its internal optimization cost (Fig.~\ref{fig_PID_LQR_MPC}(d)).

In contrast, PID controller, which lacks long-horizon feedback, shows higher sensitivity~\cite{sicilianoRobotics2009}. Fig.~\ref{fig_PID} plots the second joint's posture difference and end-effector trajectory difference over time under different quantization settings with PID controller. Although posture error is not used as a precision metric in our work, the framework includes optional support for such metrics to match application-specific needs. Fig.~\ref{fig_PID} shows that errors remain negligible during large corrections (first 10,000 steps), but accumulate noticeably during fine convergence stages. For example, 8-bit fractional quantization produces over 1mm trajectory error near the final target position.

The results show that 16-bit and 12-bit fractional configurations provide adequate precision, though motion degradation remains visible in high-precision tasks. Notably, even existing accelerators using 32-bit fixed-point (16-bit fractional) data types~\cite{DaduRBD,Roboshape} exhibit measurable effects on motion accuracy, supporting the need for precision-aware quantization tuning. This underscores the importance of our framework, demonstrating its potential to improve hardware efficiency while maintaining motion accuracy.

Based on bit-width formats searched specifically for FPGA implementation (18/24-bit DSP word sizes) to maximize hardware efficiency, we adopt quantization methods as follows: 24-bit (12 int / 12 frac) for iiwa, 18-bit (10 int / 8 frac) for HyQ, and 24-bit (12 int / 12 frac) for Atlas. Each setting is automatically derived by our framework within 8 hours on a workstation with an Intel Core i9-12900 CPU—significantly faster than~\cite{hsiaoVaPrVariablePrecisionTensors2023}. The resulting formats are then used for FPGA-based RBD accelerator design and evaluation.

\subsection{FPGA Accelerator Evaluation}

\begin{table}[t]
    \caption{Harware Configurations for Evaluations.}
    \centering
    \vspace{-6pt}
    \label{tab_harware_configurations}
    \renewcommand\arraystretch{1.06} 
    \begin{tabular}{|c|c|c|c|}
    \hline
    \textbf{Type} & \textbf{Hardware Platform} & \textbf{Freq} & \textbf{Evaluated in} \\ \hline
    CPU      & Jetson AGX Orin         & 2.2G      & \cite{AnalyticalDerivativesRSS,plancherAcceleratingRobotDynamics2021}      \\
    CPU      & Core i9-12900         & 5.1G      & \cite{AnalyticalDerivativesRSS,plancherAcceleratingRobotDynamics2021}      \\
    GPU      & Jetson AGX Orin         & 1.3G      & \cite{plancherGRiDGPUAcceleratedRigid2022}      \\
    GPU      & RTX 4090M        & 1.8G      & \cite{plancherGRiDGPUAcceleratedRigid2022}      \\
    FPGA     & XCVU9P           & 56M       & \cite{Roboshape}      \\
    FPGA     & XCVU9P  & 125M      & \cite{DaduRBD}      \\
    FPGA     & XCV80 \& U50 & 228M      & DRACO      \\ \hline
    \end{tabular}
    \vspace{-18pt}
\end{table}

\begin{figure*} [t]
    \centering
    \includegraphics[width=\linewidth]{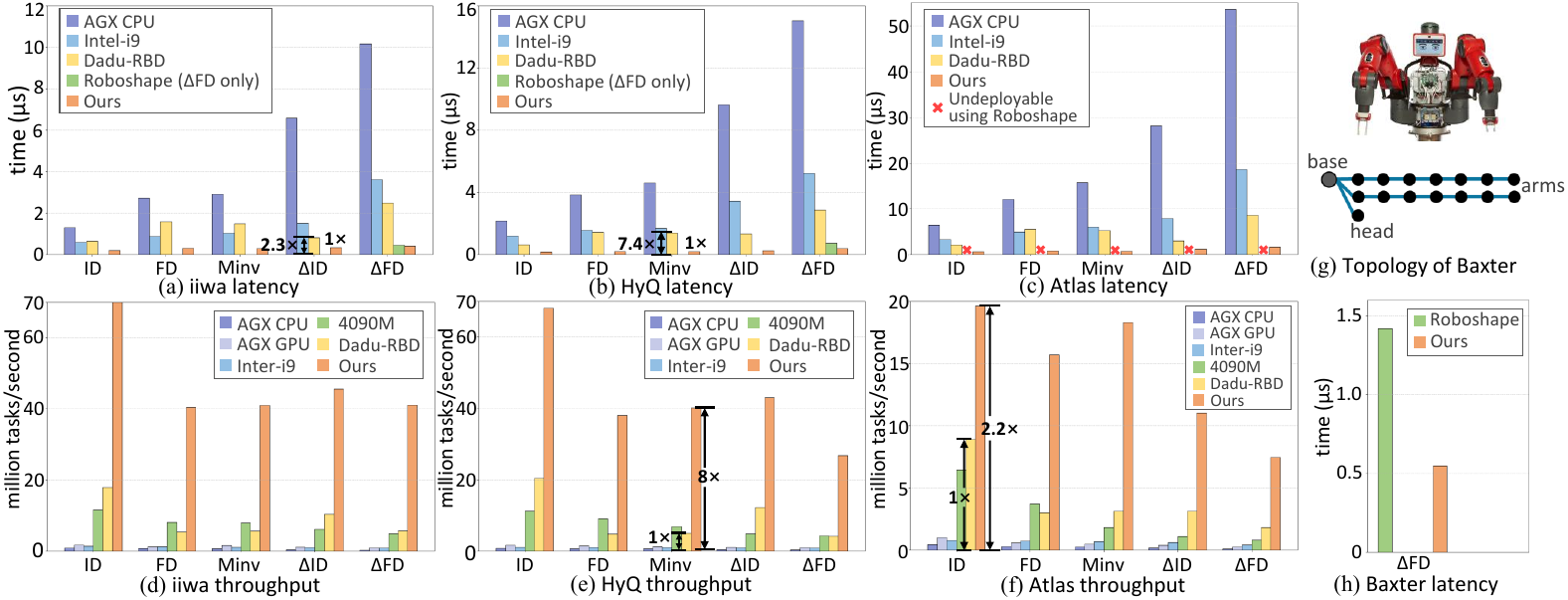}
    \vspace{-22pt}
    \caption{Performance comparisons with Pinocchio~\cite{AnalyticalDerivativesRSS} on CPU, GRiD~\cite{plancherGRiDGPUAcceleratedRigid2022} on GPU, Roboshape~\cite{Roboshape} and Dadu-RBD~\cite{DaduRBD} on FPGA.}
    \label{fig_latency_throughput}
    \vspace{-18pt}
\end{figure*}

We implement DRACO architecture on two FPGAs: an AMD Alveo V80~\cite{AMDV80} with DSP58~\cite{DSP58} (24-bit, for iiwa and Atlas) and an AMD Alveo U50~\cite{AMDU50} with DSP48~\cite{DSP48} (18-bit, for HyQ), both synthesized using Vivado 24.1 and running at 228MHz. Although these FPGAs differ from the XCVU9P FPGA used in baselines~\cite{DaduRBD,Roboshape}, the comparison remains fair: resource usage is similar and within platform limits (Table~\ref{tab_harware_resource}), and FPGA memory interface bandwidth match the configuration in Dadu-RBD~\cite{DaduRBD}. We compare DRACO against widely used baselines, including a C++-based RBD library~\cite{AnalyticalDerivativesRSS}, CPU~\cite{plancherAcceleratingRobotDynamics2021}, GPU~\cite{plancherGRiDGPUAcceleratedRigid2022}, and FPGA accelerators~\cite{Roboshape,DaduRBD}, with a focus on the SOTA FPGA design Dadu-RBD~\cite{DaduRBD}. Table~\ref{tab_harware_configurations} summarizes hardware configurations, including Jetson AGX Orin (advanced edge computing module), and i9-12900 and 4090M (near-top-tier mobile CPU/GPU). 

DRACO is evaluated on four robots: KUKA iiwa~\cite{iiwa}, HyQ~\cite{HyQDynamicLegged}, Atlas~\cite{Atlas}, and Baxter~\cite{Baxter} (Fig.\ref{fig_latency_throughput} (g)), consistent with prior studies~\cite{AnalyticalDerivativesRSS,plancherGRiDGPUAcceleratedRigid2022,DaduRBD,Roboshape}. Among existing works, only Roboshape~\cite{Roboshape} reports the latency of $\Delta$FD function for the Baxter robot. For comparison, we use 24-bit quantization method (12 int / 12 frac) and implement it on the V80 FPGA.

\textbf{Latency and Throughput.} Latency is evaluated against CPU/FPGA baselines (excluding GPUs due to their high per-task response time), and throughput is compared across CPU/GPU/FPGA baselines. Some results are taken directly from~\cite{DaduRBD}, while Roboshape’s throughput is not reported in~\cite{Roboshape}. Following the evaluation methods in~\cite{DaduRBD}, latency is measured by running 128 single-threaded tasks, and throughput is evaluated with 256 batched tasks. 

Fig.~\ref{fig_latency_throughput} shows the results. Our design uses a DSP count similar to Roboshape and slightly higher than Dadu-RBD, but remains well below FPGA’s resource capacity. Detailed comparisons of resource usage and performance per DSP will be discussed later. DRACO improves throughput over Dadu-RBD by 2.2$\times$--8$\times$, and reduces latency by 2.3$\times$--7.4$\times$. Compared with Roboshape, latency is reduced by 1.1$\times$--2.6$\times$. These improvements are primarily due to the fixed-point quantization of RBD functions. Specifically, by replacing 32-bit MACs with 24-bit or 18-bit versions, significant DSP savings allow for more parallel MAC operations, improving performance. The resource savings also enable efficient acceleration for complex robots like Atlas. As shown in Fig.~\ref{fig_latency_throughput}(c) and (f), DRACO delivers Atlas speedup comparable to simpler robots.

Additionally, the division deferring algorithm and architecture further accelerate Minv, FD, and $\Delta$FD functions. Fig.~\ref{fig_latency_throughput}(a), (d), and (e) demonstrate that Dadu-RBD often underperforms compared to CPU and GPU baselines for Minv and FD functions. In contrast, our design achieves 5.2$\times$--7.4$\times$ latency reduction and 5.9$\times$--8$\times$ throughput gains for Minv function over Dadu-RBD. Besides, the fully pipelined divider also shortens the critical path of the circuits and improves clock frequency, further enhancing performance.

\begin{figure} [t]
    \centering
    \includegraphics[width=\linewidth]{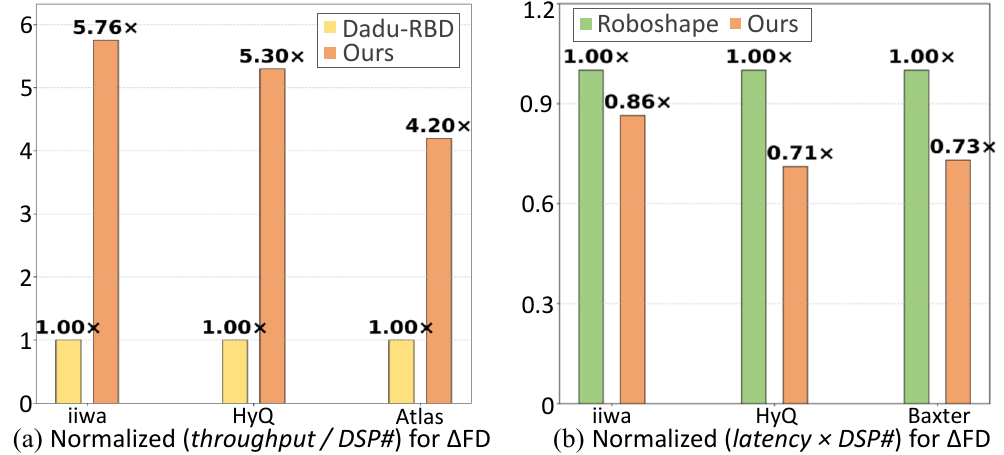}
    \vspace{-22pt}
    \caption{Normalized performance per DSP comparisons with~\cite{DaduRBD} and~\cite{Roboshape}.}
    \label{fig_throughput_per_dsp}
    \vspace{-19pt}
\end{figure}

\textbf{Performance per DSP.} We compare DRACO's performance per DSP with Dadu-RBD~\cite{DaduRBD} (SOTA throughput) and Roboshape~\cite{Roboshape} (SOTA latency). Performance per DSP is calculated as throughput divided by DSP count (higher is better) or latency times DSP count (lower is better). For iiwa, DSP usage is reported in~\cite{DaduRBD}, while for HyQ and Atlas, it is derived from our implementation of Dadu-RBD. Fig.~\ref{fig_throughput_per_dsp}(a) shows 4.2$\times$--5.8$\times$ higher throughput per DSP for $\Delta$FD under DRACO, thanks to quantization and inter-module DSP reuse.

As shown in Fig.~\ref{fig_throughput_per_dsp}(b), DRACO achieves 0.71$\times$–0.86$\times$ the latency times DSP\# product of Roboshape, reflecting different design priorities. Roboshape optimizes for minimal latency by dedicating dual cores to forward and backward propagation, consuming substantial DSPs to maximize single-task parallelism. DRACO, in contrast, distributes computations across joints to optimize multi-task throughput while maintaining competitive latency, resulting in superior overall efficiency.

\begin{figure} [t]
    \centering
    \includegraphics[width=\linewidth]{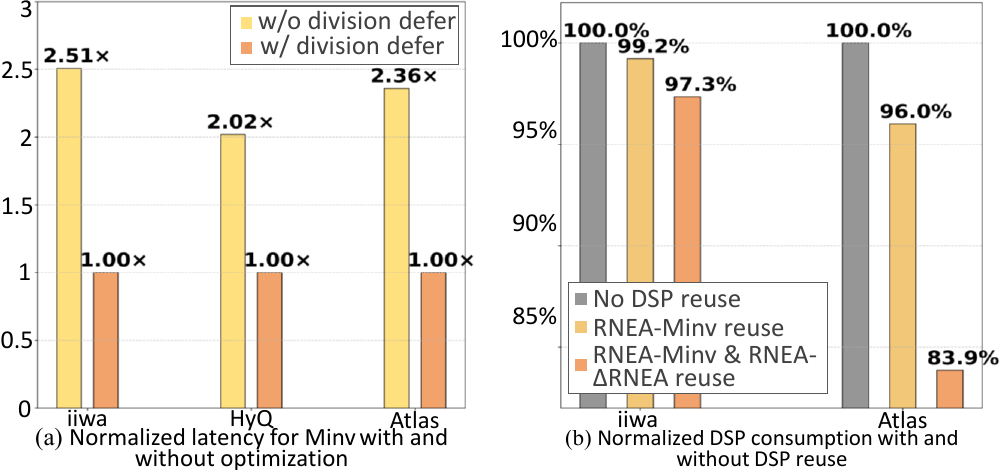}
    \vspace{-22pt}
    \caption{Normalized (a) latency for Minv w/ and w/o division deferring, and (b) DSP consumption w/ and w/o inter-module DSP reuse strategy.}
    \label{fig_Minv_latency_DSP_reuse}
    \vspace{-11pt}
\end{figure}

\textbf{Evaluation of Optimized Minv with Division Deferring.} To assess the hardware-efficient Minv algorithm and its architecture, we conduct standalone latency tests for the Minv module on the V80 FPGA with identical quantization bit-widths, DSP counts (excluding additional cost introduced by division deferring), and MAC configurations. Fig.~\ref{fig_Minv_latency_DSP_reuse}(a) shows over 2$\times$ speedup from removing reciprocal operation from the longest latency path and enabling parallel execution.

\textbf{Evaluation of Inter-Module DSP Reuse.} We compare total DSP usage before and after inter-module DSP reuse while keeping other modules unchanged. As shown in Fig.~\ref{fig_Minv_latency_DSP_reuse}(b), DSP consumption is reduced by 2.7\% for iiwa and 16.1\% for Atlas due to their differing computational demands.

For iiwa, fewer MAC operations results in smaller \textit{II} differences between RNEA, Minv, and $\Delta$RNEA modules, requiring minimal inter-module DSP reuse to balance the \textit{II}. For Atlas, with its higher computational demands, extensive DSP reuse is required within $\Delta$RNEA and Minv modules to reduce DSP consumption. But this increases \textit{II} differences between modules, as $\Delta$RNEA and Minv have heavier loads compared to RNEA. So more DSPs are reused to balance execution rates. It’s important to note that these results assume abundant DSP resources, where reuse does not degrade performance. In resource-constrained scenarios, reuse would need to be more aggressive, potentially impacting performance.

\textbf{Resource and Power Consumption.} Table~\ref{tab_harware_resource} compares resource usage across DRACO (based on implementation in Fig.~\ref{fig_latency_throughput}) and other FPGA baselines (taken from their respective papers). Beyond the tabulated LUT and DSP usage, DRACO uses 371k FFs and 167 BRAMs. In terms of power consumption, DRACO's total on-chip power for iiwa is 33.5W (9W dynamic power), comparable to 36.8W of Dadu-RBD.

\begin{table}[t]
    \caption{Hardware Resource Usage.}
    \centering
    \vspace{-6pt}
    \label{tab_harware_resource}
    \renewcommand\arraystretch{1.06} 
    \begin{tabular}{|m{1.2cm}<{\centering}|m{2cm}<{\centering}|m{1cm}<{\centering}|m{1cm}<{\centering}|m{1cm}<{\centering}|}
    \hline
    \textbf{Resource} & \textbf{Accelerator} & \textbf{iiwa} & \textbf{HyQ} & \textbf{Atlas} \\ \hline
    \multirow{3}*{DSP}   & DRACO         & 5073      & 4002  & 6301      \\ 
                           & Dadu-RBD      & 4241      & N/A   & N/A       \\ 
                           & Roboshape     & 5448      & 3008  & N/A       \\ \hline
    \multirow{3}*{LUT}   & DRACO         & 584k    & 509k  & 647k  \\ 
                           & Dadu-RBD      & 638k    & N/A     & N/A     \\ 
                           & Roboshape     & 515k    & 507k  & N/A     \\ \hline
    \end{tabular}
    \vspace{-16pt}
\end{table}

\subsection{Estimated Control Rate Improvement}

\begin{figure} [t]
    \centering
    \includegraphics[width=\linewidth]{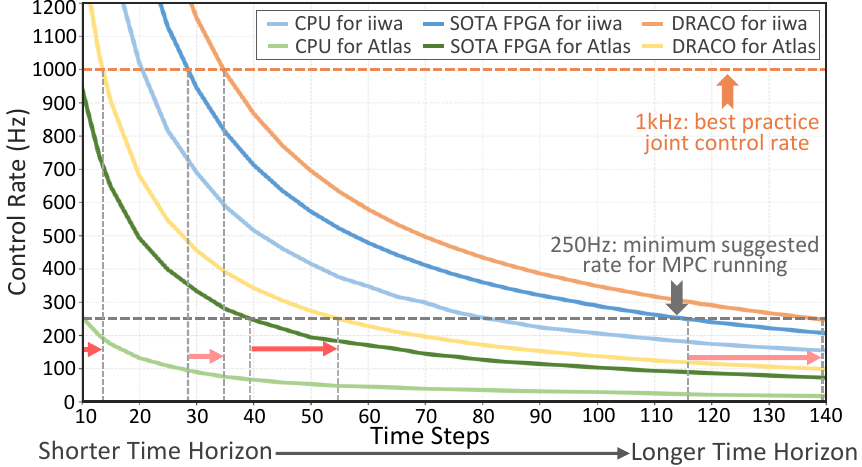}
    \vspace{-22pt}
    \caption{Estimated control rates for iiwa and Atlas based on different trajectory lengths, compared to ideal control rates required for online use (1kHz, 250Hz)~\cite{RealtimeMotionPlanningMPC}. CPU implementations are based on~\cite{EfficientAnalyticalDerivatives2022}. FPGA implementations are based on~\cite{DaduRBD}, using V80 FPGA~\cite{AMDV80}. We use the similar analytical model in~\cite{Robomorphic}, assuming 10 iterations of the MPC optimization loop.}
    \label{fig_control_rate}
    \vspace{-19pt}
\end{figure}

Using the analytical model from~\cite{Robomorphic}, we estimate the control rate improvements achieved by accelerating the RBD functions with DRACO, as shown in Fig.\ref{fig_control_rate}. DRACO enables higher control rates, allowing robots to perform more optimization loop iterations for better trajectory computation or to plan over longer time horizons. For instance, DRACO supports up to 54 time steps at 250 Hz for Atlas, compared to 39 for Dadu-RBD. To ensure a fair comparison, results of Dadu-RBD are obtained by implementing it on the V80 FPGA, which offers more hardware resources.

\section{Related Works}

\textbf{Design Methedology for RBD Accelerators.} To enable agile design as robotic applications and platforms evolve, Robomorphic~\cite{Robomorphic} and Roboshape~\cite{Roboshape} introduce automated hardware design flows that transform robot morphology into customized accelerator architectures. These methodologies leverage robot topology and structure to exploit parallelism and matrix sparsity in hardware design. However, their FPGA hardware templates are suboptimal, resulting in limited throughput and scalability. Our work addresses these limitations by introducing a high-performance, hardware-efficient accelerator template suitable for agile design.

\textbf{Accelerators for Robotic Applications.} The increasing computational complexity and diverse scenarios of robotics have led to the development of specialized accelerators for tasks such as mapping and localization~\cite{hadidiQuantifyingDesignspaceTradeoffs2021,kimSuperNoVAAlgorithmHardwareCoDesign2025,liuEnergyEfficientRuntime2023a,liuArchytasFrameworkSynthesizing2021}, motion planning~\cite{dubeHardwareSoftwareCoDesignPath2024,huangMOPEDEfficientMotion2024,lianDaduPScalableAccelerator2018,murrayMicroarchitectureRealtimeRobot2016,haoBLITZCRANKFactorGraph2023a,huangHardwareArchitectureGraph2022a,luoAcceleratingPathPlanning2022,murrayProgrammableArchitectureRobot2019,shahCollisionPredictionRobotics2024a,shahEnergyEfficientRealtimeMotion2023,sugiuraP3NetPointNetbasedPath2022,sugiuraIntegratedFPGAAccelerator2024,sunRTSARRAMTCAMBased2024,zangRobotMotionPlanning2022}, control~\cite{lianDaduAcceleratingInverse2017,sacksRoboXEndtoEndSolution2018}, and other applications~\cite{buiRealTimeHamiltonJacobiReachability2021,haoORIANNAAcceleratorGeneration2024}. Among these, DRACO specifically targets the control stage, which impose the most stringent real-time requirements.

\textbf{Quantization and Approximation for Robotic Planning and Control.} Robomorphic~\cite{Robomorphic}, Roboshape~\cite{Roboshape}, and Dadu-RBD~\cite{DaduRBD} use 32-bit fixed-point arithmetic for RBD computations but do not consider its impact on precision. VaPr~\cite{hsiaoVaPrVariablePrecisionTensors2023} applies variable precision quantization to GPU-based tensor operations in motion planning, but its method requires time-intensive, application-specific searches and lacks precise evaluations of quantization effects. Its acceleration effect is also suboptimal. Tartan~\cite{Tartan} takes an approximation-based approach, improving the performance of computing heuristic cost functions in the A* algorithm by sacrificing accuracy. However, it lacks detailed accuracy analysis and requires training a neural network for each new application scenario, making it time-consuming and inflexible. In contrast, our quantization framework automatically determines optimal bit-width configurations based on user-defined precision, range requirements, and target hardware. The ICMS component supports multiple controllers, enabling comprehensive motion-state simulations. Future enhancements may incorporate mixed-precision quantization for different RBD components to further improve resource utilization and performance.

\section{Conclusion}

We propose DRACO, an FPGA accelerator for robotics RBD functions. DRACO incorporates a RBD quantization framework that identifies optimal bit-width based on user-defined control strategies, precision and range requirements, effectively reducing hardware resource usage while maintaining accuracy. Additionally, we introduce a division deferring method to speed up Minv computations and an inter-module DSP reuse strategy to balance processing rates and enhance DSP utilization. DRACO achieves superior performance for various RBD functions across multiple robots. These advancements enable higher control rates and longer time horizons, significantly enhancing robotic system capabilities. For the ultimate goal of agilely designing RBD accelerators tailored to different robots and applications~\cite{Robomorphic,Roboshape}, our work provides a high-performance hardware template, innovative quantization framework, and a reusable DSP design pattern.


\bibliographystyle{IEEEtranS}
\bibliography{refs}

\end{document}